\documentclass[twocolumn,showpacs,preprintnumbers,showkeys,amsmath,amssymb]{revtex4}

\usepackage{hyperref}
\usepackage{graphicx}
\usepackage{color}

\begin{document}

\newcommand{\Det}{\,{\rm{Det}\,}}
\newcommand{\Tr}{\,{\rm{Tr}\,}}
\newcommand{\tr}{\,{\rm{tr}\,}}
\newcommand{\gc}{\color{blue}}
\newcommand{\rc}{\color{red}}
\renewcommand{\Im}{{\rm Im}}

\newcommand{\COMMENT}{\large\bf}
\newcommand{\FigWidth}{0.9\columnwidth}

\newcommand{\CC}{\mbox{\Bbb C}}
\newcommand{\NN}{\mbox{\Bbb N}}
\newcommand{\PP}{\mbox{\Bbb P}}        
\newcommand{\QQ}{\mbox{\Bbb Q}}
\newcommand{\RR}{\mbox{\Bbb R}}
\newcommand{\TT}{\mbox{\Bbb T}}
\newcommand{\XX}{\mbox{\Bbb X}}
\newcommand{\ZZ}{\mbox{$\mathbb Z$}}
\newcommand{\z}{\raise-1pt\hbox{$\mbox{\Bbbb Z}$}}

\def\beq{\begin{equation}}
\def\eeq{\end{equation}}
\def\p{\partial}
\def\G{\Gamma}

\def\MM{{\sf M}}
\def\DD{{\sf D}}
\def\Dc{\CC \setminus {\sf D}}
\def\Uc{{\sf U}_{{\rm ext}}}
\def\UU{{\sf U}}
\def\BB{{\sf B}}

\def\unitary{{\mathscr U}}
\def\hermitian{{\mathscr H}}
\def\normal{{\mathscr N}}
\def\complex{{\mathscr C}}

\def\kernel{K}
\def\lbracket{\left <}
\def\rbracket{\right >}
\def\vphcl{\varphi _{0}}
\def\rhocl{\rho _{0}}

\def\square{\hfill
{\vrule height6pt width6pt depth1pt} \break \vspace{.01cm}}

\title{Multi-Cut Solutions of Laplacian Growth}

\author{Ar.~Abanov}
\email[]{abanov@tamu.edu}
\affiliation{
            Department of Physics, MS 4242,
            Texas A\&M University,
            College Station, TX 77843-4242, USA
}
\author{M.~Mineev-Weinstein}
\email[]{mariner@lanl.gov} \affiliation{Los Alamos National
Laboratory,
             MS-P365,
             Los Alamos, NM 87545, USA
}
\author{A.~Zabrodin}
\email[]{zabrodin@itep.ru} \affiliation{Institute of Biochemical
Physics,
             Kosygina str. 4,
             119334 Moscow, Russia;
             also at
             ITEP,
             Bol. Cheremushkinskaya str. 25,
             117218 Moscow, Russia}

\date{\today}

\begin{abstract}

A new class of solutions to Laplacian growth (LG) with zero surface
tension is presented and shown to contain all other known solutions
as special or limiting cases.  These solutions, which are
time-dependent conformal maps with branch cuts inside the unit
circle, are governed by a nonlinear integral equation and describe
oil fjords with non-parallel walls in viscous fingering experiments
in Hele-Shaw cells. Integrals of motion for the multi-cut LG
solutions in terms of singularities of the Schwarz function are
found, and the dynamics of densities (jumps) on the cuts are derived.
The subclass of these solutions with linear Cauchy densities on the
cuts of the Schwarz function is of particular interest, because in
this case the integral equation for the conformal map becomes linear.
These solutions can also be of physical importance by representing
oil/air interfaces, which form oil fjords with a constant opening angle,
in accordance with recent experiments in a Hele-shaw cell.

\end{abstract}



\maketitle \tableofcontents

\section{Introduction}\label{sec:Introduction}

{\it Background and motivation:} The Laplacian growth (LG) is a free
boundary motion governed by gradient of a harmonic field.  It
describes numerous physical processes \cite{GH} with a moving
boundary between two immiscible phases (an interface) far from
equilibrium and profoundly interconnects various branches of
mathematical physics \cite{KMWZ,Gust}.  LG in two dimensions is of
special importance since in the zero surface tension limit it has a
remarkably rich integrable structure \cite{Richardson,MWZ,KKMWZ} and
an impressive list of exact solutions in the form of conformal maps
with pole \cite{Kuf,SB84,M90} and logarithmic
\cite{H86,BP86,MS94,SM94} time-dependent singularities (many of
these solutions cease to exist in a finite time due to the interface
instability.) In addition, there exist a family of explicit
solutions \cite{BAmar,Tu,AMZ} with time-independent branch cuts with
fixed endpoints which describe growth of self-similar air `fingers'
in a wedge geometry (or in the plane with the rotational $\ZZ
_N$-symmetry imposed).

Do other solutions exist, apart from those mentioned above?
Answering `yes' to this question, we hereby present a new family of
conformal maps with time-dependent cuts, study their dynamics and
provide examples.  This family is the most general solution to the
LG problem (in the absence of surface tension) which can be written
in terms of analytic functions.  All rational, logarithmic and
self-similar solutions known so far (and mentioned above) are shown
to be special or limiting cases of these newly found ones.

In this paper we consider a simply connected {\it exterior} 2D LG problem with a
source/sink at infinity. In the context of Hele-Shaw flows
\cite{Hele-Shaw1898} (the prototype of all 2D LG processes
\cite{RMP86}), it corresponds to a finite bubble of an inviscid
fluid (air) surrounded by a viscous fluid (oil) with a source or
sink at infinity. In what follows we hold to this hydrodynamic
interpretation.

{\it The standard formulation} of the exterior LG problem in 2D
planar geometry is as follows:
\begin{equation}
\!\!\left\{
\begin{array}{l}
\left.
\begin{array}{l}
{\bf v}  = -{\bf \nabla} p \\
\Delta p =0
\end{array}
\right\}\mbox{in }D(t)\\
\left.
\begin{array}{l}
p = \mbox{const}\\
V_n = -\p_n p \\
\end{array}
\right\}\mbox{at }\Gamma(t)\\
p(x,y)= -\frac{Q}{2\pi}\log R \mbox{, when }
R = \sqrt{x^2 + y^2} \to \infty. \\
\end{array}
\right . \label{eq:LG}
\end{equation}
Here $t$ is time, $\Gamma(t)$ is the boundary (an oil/air interface)
of an infinite planar domain $D(t)$ containing infinity and filled
with oil, $p$ is the pressure field, ${\bf v}$ is velocity of oil,
$V_n$ is normal velocity of the oil/air interface, $\p_n$ stands for
normal derivative at the interface, and $Q$ is a pumping rate
assumed to be time-independent (positive for the suction problem and
negative for the injection one).

The first equation in the system \eqref{eq:LG} is the Darcy law in
scaled units, which determines the local velocity of oil in the
Hele-Shaw cell.  The second equation follows from continuity and
incompressibility of oil, $\mbox{div}\,{\bf v}=0$.  The third
equation results from the fact that in fluids with negligible
viscosity pressure is the same everywhere in the fluid,
$p=\mbox{const}$, so, neglecting surface tension, one can set $p$ to
be the same constant along both sides of the interface. The fourth
equation states that normal velocity of fluid at the interface and
that of the interface itself coincide (continuity).  The last
equation gives the asymptotic of $p$ far away from the interface in
the presence of a single sink of strength $Q$ placed at infinity.

A reformulation of the LG problem in terms of complex variables and
analytic functions is particularly instructive.  We conclude the
introduction by a brief review of the time-dependent conformal map
approach.

{\it The conformal map formulation}. Let us introduce a
time-dependent conformal map, $z = f(w,t)$, from the exterior of the
unit circle, $|w| \geq 1$, in an auxiliary {\it mathematical
$w$-plane} to the domain $D(t)$ occupied by oil in the {\it physical
plane} $z = x + iy$. Then the interface at time $t$ is a closed
curve swept by $f(e^{i \phi},t)$ as $\phi$ runs from $0$ to $2\pi$.
Below we sometimes do not indicate the time dependence explicitly
and write simply $f(w)$. It is convenient to normalize the conformal
map by the condition that $\infty$ is mapped to $\infty$ and the
derivative at infinity is a real positive number $r$ (the conformal
radius), so that the Laurels expansion at infinity has the form
\beq
\label{eq:norm} f(w)=rw +u_0 + u_1 /w + u_2 / w^2 + \ldots
\eeq
(the coefficients $u_i$ are in general complex numbers).

It is well known \cite{Galin,PK45} that the LG dynamics of the
interface   is equivalent to the following equation for $f(e^{i
\phi},t)$:
\beq \label{eq:LGE}
\Im (\bar f_t f_{\phi}) =\frac{Q}{2\pi},
\eeq
where subscripts stand for partial derivatives
and a bar for complex conjugation. This nonlinear equation is
remarkable because it possesses an infinite number of conservation
laws \cite{Richardson} and an impressive list of non-trivial
solutions with moving singularities
\cite{Kuf,SB84,M90,H86,BP86,MS94,SM94}, which are either poles or
logarithmic branch points lying strictly inside the unit circle. In
all these cases the conformal map $f(w)$ thus admits an analytic
continuation across the unit circle, so the function $f(w)$ is
actually analytic not only in its exterior but in some larger
infinite domain containing it.  Assuming that $f(w,t)$ is
analytically extendable across the unit circle for all $t$ in some
time interval, one can analytically continue the LG equation
\eqref{eq:LGE} itself.  Set $\bar f (1/w,t)\equiv \overline{f(1/\bar
w,t)}$, then equation \eqref{eq:LGE} can be rewritten as
\beq
\label{eq:LGE1} \p_w f(w,t) \p_t \bar f(1/w,t)- \p_t
f(w,t) \p_w \bar f(1/w,t) = 1/w
\eeq
(we have set $Q=\pi$ that
means a resealing of time units).  Under our assumption the
functions $f(w,t)$ and $\bar f (1/w,t)$ have a common domain of
analyticity containing the unit circle $|w|=1$, and $w$ in equation
\eqref{eq:LGE1} is supposed to belong to this domain. In fact, for
rational and logarithmic solutions, this equation holds everywhere
in the $w$-plane except for poles and branch cuts of $f(w,t)$ and
$\bar f (1/w,t)$.  We note in advance that the general multi-cut
solutions constructed below have the same property.  Namely, the
functions $f(w,t)$ and $\bar f (1/w,t)$ have a common domain of
analyticity where they actually solve the analytically continued LG
equation \eqref{eq:LGE1} and not just \eqref{eq:LGE}.

As said earlier, there are also self-similar $\ZZ _N$-symmetric
solutions \cite{BAmar,Tu,Gust,AMZ} containing a time-independent
fractional power singularity which corresponds to the interface
self-intersection point at the origin. In this case the analytic
continuation is possible through every point on the unit circle
except those that are mapped to the origin.  Again, the functions
$f(w,t)$ and $\bar f (1/w,t)$ have a non-empty common domain of
analyticity and equation \eqref{eq:LGE1} is valid everywhere in this
domain.

\section{The Schwarz Function and Integrals of Motion}
\label{sec:Schwarz}

It is known \cite{Richardson,H86,H92,MS94,KMWZ,MWZ} that the partial
differential equation \eqref{eq:LGE1} can be integrated in terms of
the {\it Schwarz function}.  It is an analytic function, $S$, which
connects $f(w)$ and $\bar f (1/w)$ in their common domain of
analyticity: $\bar f(1/w)= S(f(w))$. Equivalently, the Schwarz
function for a curve $\Gamma$ is an analytic function $S(z)$ such
that $\bar z  = S(z)$ for $z\in \Gamma$ \cite{Davis}.  In other
words, $S(z)$ is an analytic continuation of the function $\bar z$
away from the curve.  For analytic curves this function is known to
be well defined in a strip-like neighborhood of the curve.  If the
curve depends on time, so does its Schwarz function, $S=S(z,t)$.

In terms of the Schwarz function the (analytically continued) LG
equation \eqref{eq:LGE1} reads \cite{H92}
\beq\label{eq:Sdot}
S_t =\p_z \log w
\eeq
where $w=w(z,t)$ is the function inverse to the
$f(w,t)$.  Since $\p_z \log w(z)$ is analytic everywhere in the oil
domain $D(t)$, both sides of equation \eqref{eq:Sdot} are free of
singularities there.  Therefore, all singularities of $S(z,t)$
located in $D(t)$ are time-independent. Equivalently, the function
\beq
S_+(z) = \oint_{\Gamma(t)}\,\frac{S(\zeta)}{\zeta -z}\,
\frac{d\,\zeta}{2\pi i} \label{eq:S+}
\eeq
defined for $z$ outside
$D(t)$ by the integral of Cauchy type (where the integration contour
$\Gamma$ has the standard anti-clockwise orientation) and
analytically continued to $D(t)$ is constant in time: $\p_t S_+
(z,t) = 0$.  This implies conservation of {\it harmonic moments}
$t_k$ \cite{Richardson} which are coefficients of the Taylor
expansion of $S_+(z)$ at $z=0$:
\begin{eqnarray}
&&t_k = \frac{1}{2\pi ik}\oint_{\Gamma (t)}
\frac{\bar z dz}{z^k}= \nonumber\\
&&-\frac{1}{\pi k}
\int_{D(t)}\frac{dx\,dy}{z^k},\qquad  k=1,2,3, \dots,
\label{eq:harmonicMoments}
\end{eqnarray}
where we assume, without loss of
generality, that the contour $\Gamma (t)$ encircles the origin. In
particular, $t_1 = S_+ (0)$.  Besides, it is easy to see that the
physical time $t$ is equal to the area of the air bubble divided by
$\pi$: \beq \label{eq:area} t=\oint_{\Gamma (t)}\frac{\bar z\,
dz}{2\pi i} \eeq Thus the LG dynamics is a process such that the
area surrounded by the interface grows linearly with time while the
harmonic moments \eqref{eq:harmonicMoments} (or the $S_{+}$-part of
the Schwarz function \eqref{eq:S+}) are kept constant.  It is
important to note that the function $S_+$ (i.e., the infinite set of
harmonic moments $t_k$) together with the variable $t$ fix the whole
Schwarz function $S$ uniquely, at least locally in the variety of
closed analytic contours in the plane.  In a certain sense (made
more precise in \cite{KMZ05}), these data serve as local coordinates
in the infinite dimensional variety of contours.

It is also worthwhile to mention that the function
$S_-(z)=S(z)-S_+(z)$ is analytic in $D(t)$ and vanishes at infinity
as $O(1/z)$ with the residue \beq \label{eq:resS-}
\mbox{res}_{\infty}\, [S_-(z)dz]=-t \eeq (the residue at infinity is
defined as $\mbox{res}_{\infty}[dz/z]=-1$).  The decomposition of
the Schwarz function $S=S_+ + S_-$ appears to be very useful.  As it
follows from properties of Cauchy-type integrals, the function
$S_-(z)$ is given by the same integral \eqref{eq:S+} (with the
opposite sign), where $z$ now belongs to $D(t)$.

The conformal map and the Schwarz function are connected in the
following way:
\begin{equation}
\label{eq:fS} \left\{
\begin{array}{l}
S (z) = \bar f(1/w) \\
z = f(w)
\end{array}
\right.
\end{equation}
These formulas make it clear that if the functions $f(w)$ and $\bar
f(1/w)$  have a common domain of analyticity containing the unit
circle, then the Schwarz function is well-defined. They also imply
the one-to-one correspondence between singularities of the function
$f(w)$ inside the unit disk and singularities of the function $S(z)$
in $D(t)$ (which, by the properties of the Cauchy type integrals,
are the same as singularities of the function $S_+(z)$).

Let us illustrate the one-to-one correspondence between
singularities of $f(w)$ and $S_+(z)$ by the example of poles in
which case it is especially transparent.  Suppose that the conformal
map $f(w)$ has a pole of order $k$ at a point $a$ inside the unit
disk, so that the local behavior of $f$ near $a$ is \beq f (w)=
\frac{A}{(w-a)^k} \,\,\, + \,\,\, \mbox{less singular terms}.
\label{eq:pole} \eeq Then the function $S = \bar f(1/w)$ has a pole
of the same order at the point $w = 1/\bar a$: \beq \label{eq:pole1}
S = \bar f(1/w)  = \frac {\bar A}{(w^{-1} - \bar a)^k} \,\,\, +
\,\,\, \mbox{less singular terms}. \eeq Because the point $1/\bar a$
together with its small neighborhood lies outside the unit disk
(where $f$ is conformal), we can linearize the denominator in
\eqref{eq:pole1} near $z=f(1/\bar a)$ and obtain
\begin{equation}
S(z) =  \bar A \left[\frac{f_{\bar a}(1/\bar a)}{z - f(1/\bar a)}
\right]^k \,\,\, + \,\,\, \mbox{less singular terms}
\label{eq:poleS}
\end{equation}
(here $f_{\bar a}(1/\bar a)\equiv \p_w f(1/w)$ at $w=\bar a$).  Thus
$S(z)$ necessarily has a singularity at $z = f(1/\bar a)$ of the
same type as $f(w)$ has at $w = a$.

Now we are ready to outline the general strategy of integration of
the LG problem. Taken initial data, i.e., a conformal map $f(w)$ at
$t=t_0$, one finds $S_+$ from \eqref{eq:S+} by rewriting it as
integral over the unit circle in the mathematical plane:
\beq\label{eq:S+1}
S_+(z)=\frac{1}{2\pi i}\oint_{|w|=1} \frac{\bar
f(1/w)\, d f(w)}{f(w)-z}\, .
\eeq
In order to obtain the LG
dynamics, one then should solve the inverse problem: given the
$S_+(z)$ and time $t$, to recover the conformal map $f(w,t)$.  The
last step is a part of the inverse potential problem and thus is
hard to implement in general.  Formally, this is equivalent to a
nonlinear integral equation which is easy to derive using the
definition of the Schwarz function in the form $f(\zeta )=\bar
S(\bar f(1/\zeta ))$. Integrating both sides of this equality with
the Cauchy kernel $1/(w-\zeta )$ over the unit circle assuming that
$|w|>1$, and using the fact that the $S_-$-part of the Schwarz
function does not contribute to the integral, we get the integral
equation
\beq \label{eq:int1}
f(w)=rw+u_0 +\frac{1}{2\pi
i}\oint_{|\zeta |=1}\!\!\!\!\!\!\!\! \frac{\bar S_{+}(\bar f (1/\zeta
))d\zeta}{w-\zeta}, \quad |w|>1.
\eeq
The same procedure at
$|w|<1$ gives a ``complimentary" equation
\beq \label{eq:int1bis}
rw+u_0 =\frac{1}{2\pi i}\oint_{|\zeta |=1}\!\!\!\!\!\!\!\!
 \frac{\bar S_{+}(\bar f
(1/\zeta ))d\zeta}{\zeta -w} + \bar S_{-}(\bar f(1/w)), \,
|w|<1.
\eeq
As their direct consequence, we can write the
following useful formulae for the coefficients of the Laurels series
\eqref{eq:norm}: \beq \label{eq:int2} u_j=\frac{1}{2\pi
i}\oint_{|\zeta |=1} \bar S_{+}(\bar f (1/\zeta ))\zeta ^{j-1}
d\zeta \,, \quad j\geq 0\,, \eeq \beq \label{eq:int3}
r=\frac{1}{2\pi i}\oint_{|\zeta |=1} \bar S_{+}(\bar f (1/\zeta
))\zeta ^{-2} d\zeta + \frac{t}{r} \eeq Substituting them back into
\eqref{eq:norm} and using \eqref{eq:int1}, we get a set of integral
equations
\begin{eqnarray} \label{eq:int4a}
&&f(w)=rw + \sum_{j=0}^{d}u_j w^{-j} +\nonumber\\
&&\frac{1}{2\pi i}\oint_{|\zeta |=1} \frac{\zeta ^{d} \bar S_{+}(\bar
f (1/\zeta )) d\zeta}{w^d (w-\zeta )}, \qquad |w|>1.
\end{eqnarray}
for any
$d=-1, 0, 1, 2, \ldots$.  Being taken together with \eqref{eq:int2},
all of them are equivalent to the original equation \eqref{eq:int1}
which is reproduced at $d=0$. However, depending on the type of
singularities of the function $S_+$, one or another form may be more
convenient than others.  The choice $d=-1$ gives an equation where
none of the coefficients $u_j$ enter explicitly:
\beq
\label{eq:int4}
f(w)=rw +\frac{1}{2\pi i}\oint_{|\zeta |=1}
\frac{w\bar S_{+}(\bar f (1/\zeta ))d\zeta}{\zeta (w-\zeta )}\,,
 |w|>1.
\eeq
This equation combined with relation
\eqref{eq:int3} accomplishes integration of the LG problem.  Let
$f(w)=f(w|r)$ be a solution to equation \eqref{eq:int4} depending on
the parameter $r$, then the time dependence is found from
\eqref{eq:int3} which determines $r$ as an implicit function of $t$:
\beq \label{eq:int5} t=r^2 -\frac{r}{2\pi i}\oint_{|\zeta |=1} \bar
S_{+}(\bar f (\zeta |r))d\zeta \eeq Alternatively, one may use the
relation \beq \label{eq:area1} t=\frac{1}{2\pi i} \oint_{|w|=1}\bar
f(1/w)\, d f(w) \eeq leading to the same results.

In principle, this scheme provides a general solution to the LG
problem but in a rather implicit form. However, for several
important classes of functions $S_+(z)$ it can be made more
explicit. In all effectively solvable cases, the r.h.s. of the
integral equation simplifies drastically after shrinking the
integration contour to singularities of the function $\bar
S_{+}(\bar f (1/\zeta ))$ inside the unit disk.  Some examples are
given in the next section.

\section{Rational and Logarithmic Solutions}
\label{sec:polsAndLogs}

In this section we apply the general method
based on equations \eqref{eq:int4}, \eqref{eq:int5}
to construction of LG solutions
with poles and logarithmic singularities.

\subsection{Rational Solutions}
\label{sec:pols}

By rational solutions we mean solutions to equation \eqref{eq:LGE1} whose only
singularities are poles inside the unit disk (not mentioning a simple pole
at infinity). The very fact that the rational ansate
is consistent with the LG equation, i.e., that the number of poles
is conserved and singularities of other types are not generated, is a
consequence of the integral equation \eqref{eq:int1}.

Let us take $S_+$ to be a rational function of a general form
\beq
S_+(z) =  \sum_{m=1}^N\,\sum_{k =1}^{K_m}
\frac{T_{m,k}}{(z-z_m)^k} + \, T_{0,0} + \, \sum_{k=1}^{K_0}T_{0,k}z^k
\label{eq:S+poles}
\eeq
where all the poles $z_m$ are in $D(t)$ and $T_{m,k}$
are arbitrary complex constants. Then the function
$\bar S_{+}(\bar f (1/\zeta ))$ has a pole at $0$
of order $K_0$ and poles of orders $K_m$
at the points $a_m$ such that
$z_m = f(1/ \bar a_m )$ and does not have other singularities
inside the unit disk. For convenience, we set $a_0 =0$. Therefore,
the integral in the r.h.s. of \eqref{eq:int1} is equal to the sum
of residues at these poles. Calculating the residues, we obtain
an expression for the conformal map in the form of a rational
function of $w$ with poles
of orders not higher than
$K_m$ at the points $a_m$ inside the unit disk:
\beq
f(w,t) = r(t) w + u_0(t)+
\sum_{m=0}^N\,\sum_{k=1}^{K_m}\frac{A_{m,k}(t)}{(w-a_m (t))^k},
\label{eq:poleF}
\eeq
The pole at $a_0 =0$
is distinguished among the others because it does not move.
All other poles as well as all coefficients depend on time.
The coefficients $A_{m,k}$ are expressed through derivatives
of the function $\bar f(1/w)$ at $w=a_m$. The general
formulae are quite complicated. In fact what we really need
are inverse formulae which express the integrals of motion
$T_{m,k}$ through time-dependent coefficients of $f(w,t)$.
Obviously,
$$
T_{m,k}=\mbox{res}_{z_m} \left [ (z-z_m)^{k-1} S_{+}(z)
dz \right ]\,, \quad m\neq 0\,,
$$
and the function $S_{+}(z)$ here can be substituted for the
$S(z)$ because $S_-(z)$ is regular in $D(t)$. Passing then to
the mathematical plane, we obtain the full list of constants
of motion:
\begin{eqnarray} \label{eq:poleZ}
&&z_m \, =  f(1/\bar a_m (t) , t)\,, \quad m\neq 0\\
&&T_{m\not=0,k} = \mbox{res}_{\bar a^{-\!1}_m \!(t)}
\left [ (f(w,t)\! -\! z_{m} )^{k-1}\bar f(w^{-1}\!\!,t)\nonumber
d f(w,t)\right ]
\\
&&T_{0,k}\, = -\, \mbox{res}_{\infty}
\left [ (f(w,t))^{-k-1}\bar f(1/w,t)
d f(w,t)\right ].\nonumber
\end{eqnarray}
We note that $T_{m,k}$
have the meaning of ``Whitham times" for a general
Whitham hierarchy which covers the LG problem with
zero surface tension. Each Whitham time $T_{m,k}$ is
``coupled" to its own type of singularity.
From this point of view,
formulae \eqref{eq:poleZ} are relations of the
hodograph type which provide a solution of the full
genus zero Whitham hierarchy in an
implicit form \cite{Krichever94}.

Algebraic equations \eqref{eq:poleZ} express time-dependent parameters
$r(t)$, $u_0(t)$, $a_m(t)$ and $A_{m,k}(t)$
implicitly via the constants of motion
$z_m$ and $T_{m,k}$. To obtain a complete set
of relations, we also need the equation
\eqref{eq:area1} containing time $t$ explicitly.
In our case the integral
is reduced to a sum of residues:
$$
t=-\mbox{res}_{\infty} [\bar f(w^{-1},t)d f(w,t)]
-\sum_{m}\mbox{res}_{\bar a^{-\!1}_m (t)}
[\bar f(w^{-1},t)d f(w,t)]
$$
Comparing with \eqref{eq:poleZ} at $k=1$, we
see that the residues at $1/\bar a_m$ are just constants $T_{m,1}$.
Therefore,
\beq \label{trational}
t=-\mbox{res}_{\infty} [\bar f(1/w,t)d f(w,t)] -\sum_m T_{m,1}.
\eeq
So, we have $N + K+1$ complex algebraic equations, where
$K=\sum_{m=0}^{N}K_m$, and one real algebraic equation for $N+K+1$
complex parameters and one real parameter which determine them as
functions of time and $N+K+1$ constants of motion. Given initial
conditions in the form \eqref{eq:S+poles},
a solution to this system
allows one to recover the time-dependent
conformal map, according to the
strategy outlined at the end of Section \ref{sec:Schwarz}.

In some important particular cases the general expressions
given above become simpler and
more explicit. For example, if none of the
poles $a_m$ lies at the origin (i.e., $A_{0,k}=0$ for all $k\geq 1$),
then $T_{0,k}=0$ for $k\geq 1$, $T_{0,0}=f(0,t)$ and
\beq\label{eq:area2}
t=r(t)\bar f'(0,t)-\sum_m T_{m,1},
\eeq
(here $f'(0,t)$ is the derivative of $f(w,t)$ at
$w=0$).
If, moreover, all poles in
\eqref{eq:poleF} are simple and none of them
lies at the origin ($K_m=1$ for $m\neq 0$ and
$K_0 =0$), then
the expressions for $T_{m,1}$ and $t$ acquire an
especially compact and
explicit form:
\begin{eqnarray}
&&T_{m,1}
= \left.\bar A_{m,1}(t)\p_w f(1/w,t)\right |_{w=\bar a_m (t)}=\nonumber\\
&&\bar A_{m,1}(t)\, \left (-\frac{r(t)}{\bar a_{m}^{2}(t)}
 + \sum_{l}\frac{A_{l,1}(t)}{(1- a_l (t)\bar a_m (t))^{2}} \right ),
\label{eq:Tm1}\\
&&
t= r(t)\bar f'(0,t)-\sum_m T_{m,1} =\nonumber\\
&&r^2 (t) + \sum_{l,m}\frac{A_{l,1} (t)
\bar A_{m,1} (t)}{(1- a_l (t)\bar a_m (t))^2}.\label{eq:area3}
\end{eqnarray}

\subsection{Logarithmic Solutions}
\label{sec:Logs}

For the class of logarithmic solutions, the function $S_+$
is taken in the form
\beq \label{eq:logS+}
S_{+}(z)=t_1 + \sum_{m=1}^{N}\bar A_m \log
\left (1- \frac{z}{z_m} \right  )\,,
\eeq
where all the branch points $z_m$ are in $D(t)$ and
$t_1$, $A_m$ are arbitrary complex constants (note that
this $t_1$ is the first harmonic moment from
\eqref{eq:harmonicMoments}). If
$$
A_0 = -\sum_{m=1}^{N} A_m  \, \neq 0\,,
$$
then there is also a branching at infinity.
This function is multi-valued and one should fix a
single-valued branch.
In a neighborhood of the origin, we
define it by the condition that $S_{+}(0)=t_1$.
In order to continue it unambiguously to $D(t)$, it is necessary
to introduce a system of cuts connecting all the branch points
in such a way that the domain $D(t)\setminus \{\mbox{all cuts}\}$
be simply connected. There are many ways to draw the cuts.
Although all of them are ultimately
equivalent, it would be convenient for us to
meet some requirements natural for the growth problem:
to respect the democracy
among the branch points $z_m$ and to ensure
that the cuts always remain in $D(t)$. For the former, let us fix
a point $q$ in $D(t)$ and make cuts from $q$ to all $z_m$ (and,
if necessary, to infinity). For the latter,
it is natural to choose $q=\infty$ since $\infty$ is the only point
in $D(t)$ which remains there forever irrespectively of
initial conditions. So, we fix a system of (non-intersecting)
cuts from the points $z_m$ to $\infty$. Note that this choice of
cuts implies that $S_+(z)$ does not have a definite value at $z=\infty$,
even if $\infty$ is a regular point, because the limit
depends on a particular way of tending $z\to \infty$.

Reconstructing $f$ from \eqref{eq:int1} in a similar way
as for rational solutions, we introduce the points $a_m$
such that $f(1/\bar a_m )=z_m$ inside the unit disk and
notice that the integral in
\eqref{eq:int1} is shrunk to residues of the differential
$$
d\bar S_{+}(\bar f (1/\zeta ))=
\sum_{m=1}^{N} A_m d\log (\bar f (1/\zeta )-\bar f (1/a_m ))
$$
which are easy to calculate.
In this way one obtains
\begin{eqnarray}
&&\!\!\!\!\!\!\!\!\!\!\!f(w,t) = r(t) w + u_0(t)+
\sum_{m=0}^{N} A_{m}\log (w-a_m (t)), \label{eq:logF}\\
&&\!\!\!\!\!\!\!\!\!\!\!\sum_{m=0}^{N} A_m =0.\nonumber
\end{eqnarray}
Again, we distinguish a possible
branch point at the origin and set $a_0 =0$. By construction,
the cuts of the function $f(w)$ go from $0$ to $a_m$ inside the
unit disk. Similarly to the Schwarz function, $f(w)$ with this choice
of cuts does not have a definite value at $w=0$,
even if $0$ is a regular point, because the limit
depends on a particular way of tending $w\to 0$.

The constants of motion are given by
\begin{eqnarray}
&&t_1 =
\frac{1}{2\pi i}\oint_{|w|=1}\frac{\bar f(1/w)d f(w)}{f(w)}= \nonumber\\
&&\bar u_0 +\sum_{m=1}^{N}\bar A_m \log \left (
\frac{\bar a_m z_m}{r}\right )\,,\nonumber
\\
&&z_m \, =  f(1/\bar a_m (t) , t)= \nonumber\\
&&r/\bar a_m (t) +u(t)+ \sum_{l=1}^{N} A_l \log (1-a_l(t)\bar a_m (t)),
\label{eq:logS+1}
\end{eqnarray}
in agreement with previous
results \cite{H86,BP86,MS94,SM94}.
Formulae \eqref{eq:int5} and
\eqref{eq:area1} allow one to derive few differently looking
equivalent expressions for $t$. The simplest one reads
\beq \label{eq:logareabis}
t=r^2 +\sum_{m=1}^{N} A_m \left (\bar z_m -\bar u_0 -ra_{m}^{-1}\right )\,.
\eeq
The others can be also useful in some situations:
\begin{eqnarray}
&&\!\!\!\!\!\!\!\!\!\!\!t=\left.
r\p_w (\bar f(w)+\bar A_0 \log f(1/w))\right |_{w=0}
+\sum_m \bar A_m z_m,\label{eq:logarea}\\
&&\!\!\!\!\!\!\!\!\!\!\!t=r^2 +\sum_{l,m}A_l \bar A_m \log (1-a_l \bar a_m).
\label{eq:logarea0}
\end{eqnarray}
The solution contains $N+1$
complex ($a_m$ and $u_0$) and one real
($r$) time-dependent parameters. They are to be
determined from the system of
$N+1$ complex equations \eqref{eq:logS+1}
and one real equation \eqref{eq:logarea}.

\subsection{``Mixed" Rational-Logarithmic Solutions}
\label{sec:Mixed}

In a similar way, one can also
consider ``mixed" solutions with both poles and logarithms.
In this case the function $S_+$ is
\begin{eqnarray}
&&S_+(z) =  \sum_{m=1}^{N}\,\sum_{k =1}^{K_m}
\frac{T_{m,k}}{(z-z_m)^k}+ T_{0,0} + \nonumber\\
&&\sum_{k=1}^{K_0}
T_{0,k}z^k +
\sum_{m=1}^{N}\bar A_m \log \left (1-\frac{z}{z_m} \right )
\label{eq:S+mixed}
\end{eqnarray}
where a single-valued branch is fixed by the condition that
\beq \label{eq:single}
S_+(0)=\sum_{m=1}^{N}\sum_{k=1}^{K_m}\frac{T_{m,k}}{z_{m}^{k}} +T_{0,0} =t_1\,.
\eeq
The equation \eqref{eq:int1} gives
\begin{eqnarray}
&&f(w,t) = r(t) w + u_0(t)+
\sum_{m=0}^{N}\,\sum_{k=1}^{K_m}\frac{A_{m,k}(t)}{(w-a_m (t))^k}+\nonumber\\
&&\sum_{m=0}^{N} A_{m}\log (w-a_m (t))\,, \quad \sum_{m=0}^{N} A_m =0\,,
\label{eq:mixedF}
\end{eqnarray}
with the same convention $a_0 =0$.
The constants of motion $z_m$,
$T_{m,k}$ are expressed through the
time-dependent parameters of the conformal map
by means of formulas similar to \eqref{eq:poleZ}.
In fact one can represent the
``hodograph relations" \eqref{eq:poleZ}
in a form
which is suitable for logarithmic and mixed cases as well:
\begin{eqnarray} &&z_m \, =  f(1/\bar a_m (t) , t)  \nonumber\\
&&T_{m\not=0,k}\, =\displaystyle{ -\, \frac{1}{k}
\mbox{res}_{1/\bar a_m (t)}
\left [ (f(w,t)-z_{m})^{k}d\bar f(1/w,t) \right ]},
\nonumber\\
&&T_{0,k\not=0}\, =\displaystyle{ -\, \frac{1}{k}\mbox{res}_{\infty}
\left [ (f(w,t))^{-k}d\bar f(1/w,t)\right ]},
\label{eq:mixedZ}
\end{eqnarray}
The residue at infinity is well defined since
the differential $d\bar f(1/w,t)$ is single-valued.
Note that the coefficient $\bar A_m$ can be formally
understood as $T_{m,0}$: $\bar A_m = -kT_{m,k}$ at $k=0$.
An expression for $T_{0,0}$ follows from \eqref{eq:single}
and the integral
formula for $t_1$ in \eqref{eq:logS+1} valid in all cases:
\beq\label{eq:T00}
T_{0,0}=\frac{1}{2\pi i}\oint_{|w|=1}\frac{\bar f(1/w)d f(w)}{f(w)}
-\sum_{m=1}^{N}\sum_{k=1}^{K_m}\frac{T_{m,k}}{z_{m}^{k}}\,.
\eeq
Different equivalent versions of the formula for $t$ can be
obtained from \eqref{eq:int5} or \eqref{eq:area1}.

Finally, we note that rational
terms in $f(w)$ can be regarded
as a special (singular) limiting case of logarithmic ones
with merging branch points.
For instance,
$$
\lim_{a_1 \rightarrow a} \left (rw + A\log\frac{w-a_1}{w-a}
\right ) = rw +
\frac{\alpha}{w-a}$$
with $ A =\alpha/(a-a_1)$.
All logarithmic, rational and ``mixed" solutions
belong to the same class characterized by the property that
the first derivative of $S_+$ is a rational function (or,
equivalently, the first derivative of $S$ is a mesomorphic
function in $D(t)$).

\section{Multi-Cut Solutions with Analytic Cauchy Densities of General Type}
\label{sec:multi}

\subsection{Motivation}
\label{sec:remarks}

It is known that all rational solutions
(except for those of the form $f(w,t) = r(t) w + u(t) + A(t)/w$,
which describe a self-similar growth of ellipse) cease to exist
in a finite time, because the dynamics gives rise to a cusp-like
singularity of the interface.  These physically meaningless singularities
just signify that the surface tension effects can not be neglected in a
vicinity of highly curved parts of the interface. Nevertheless, a
considerable subclass of the purely logarithmic solutions is well
defined for all positive times and describes a non-singular interface
dynamics at zero surface tension \cite{H86,BP86,MS94,SM94,SM98}.
Furthermore, each logarithmic term in \eqref{eq:logF} has a clear
geometric interpretation of a fjord of oil {\it with parallel walls}
left behind the advancing interface. Its vertex (the stagnation point)
is located at $z_k - A_k \log 2$, the width is $\pi|A_k|$, and the angle
between its central line and the real axis in the physical plane is
$\arg A_k$ (for more details see \cite{MS94,SM94,SM98}).  This
interpretation was found to be in an excellent agreement with
some experiments (see FIG.2 in \cite{Paterson} and numerical
work \cite{Sander}).  However, more often than not, fjords of oil left behind
the moving fronts have {\it non-parallel walls}.  Moreover,
their walls are not
always straight, but more often curved, and a non-zero opening
angle along the fjord was observed \cite{Couder,Leif}.
Such shapes of fjords
can not be explained by conformal maps with a finite number of
logarithmic or rational terms.
This was a significant motivation for us to search
for a more general class of LG solutions. From mathematical
point of view, it also looks quite natural to extend the method
developed above to
solutions with singularities of more general type than just
poles or logarithmic branch points.

Below we apply the strategy outlined at the end of Section
\ref{sec:Schwarz} to derive a closed
set of equations for the case when
$S_{+}(z)$ has branch cuts of general type
with analytic Cauchy densities outside the interface.
We start from a multi-cut ansatz for $S_{+}(z)$ and then
derive an integral equation for the conformal map $f(w)$.

\subsection{A Multi-Cut Ansatz for $S_{+}(z)$}
\label{sec:branch}

The function $S_{+}(z,t)$ for $z$ in $D(t)$ is
defined as an analytic continuation
from the air domain, where it is given by the
integral of Cauchy type \eqref{eq:S+}.
In the process of analytic continuation one necessarily encounters
singularities.
Typical singularities are branch points and poles.
In order to keep the function $S_{+}(z)$ single-valued
one has to introduce a system of branch cuts.
The analytic continuation is achieved by a deformation
of the integration contour. Moving it towards infinity as far
as possible, we can write:
\begin{eqnarray}
&&S_{+}(z)=S_{+}(0) + \sum_{{\footnotesize {\rm cuts}}} \int
\frac{z[S_+ (\tau )]_{\footnotesize {\rm cut}}}{\tau (\tau -z)}\,
\frac{d\tau}{2\pi i} - \nonumber\\
&&\sum_{{\footnotesize {\rm poles}}}\mbox{res}
\left (\frac{z S_+ (\tau )d\tau}{\tau (\tau -z)}\right )
+\oint_{|\tau |=R, R \to \infty}
\frac{zS_+ (\tau )}{\tau (\tau -z)}\, \frac{d\tau}{2\pi i}\nonumber
\end{eqnarray}
where $[S_+ (\tau )]_{\footnotesize {\rm cut}}$
denotes a discontinuity (a jump) of the
function $[S_+ (\tau )]$ across a branch cut.
Let us
denote branch points by $z_m$, by analogy with logarithmic branch
points from the previous section.
As in that case, and for the same reasons,
it is convenient to make cuts $\Gamma_m$
from $\infty$ to $z_m$.
As before, we assume that
the number of branch points is finite.
Assuming also that
$S_{+}(z)$ does not have other singularities
(in particular, poles) in
$D(t)$, a single-valued branch of this function
can be represented as a sum of Cauchy type integrals
with Cauchy densities
$\mathcal{P}_{m}(\tau )$
along the cuts plus a complex constant
$t_1 =S_{+}(0)$:
\begin{equation}\label{eq:Scuts}
S_{+}(z)=t_1 +
\sum_{m}\int_{\infty ,\Gamma_{m}}^{z_{m}}
\frac{z\mathcal{P}_{m}(\tau )}{\tau (\tau -z)}\frac{d\tau }{2\pi i}\,.
\end{equation}
The ansate \eqref{eq:Scuts} implies that the jump of
$S_+$ on the cut $\Gamma_{m}$ is
\beq \label{eq:jump1}
[S_+(z)]=S_+(z_+)-S_+(z_-)=\mathcal{P}_{m}(z)\,,
\quad z\in \Gamma_m
\eeq
where $z_+$ (respectively, $z_-$) tends to the point
$z\in \Gamma_m$ from the left (respectively, from the right)
side of the cut oriented
from the lower limit of integration to the upper one.

There is
a big freedom in how to draw the cuts.
However, if the Cauchy densities can be
analytically continued from the cuts, then
different choices of the cuts with the same
endpoints are equivalent.
Indeed, consider another system of cuts, $\tilde{\Gamma}_{m}$,
connecting the same points.
Let $\tilde{\mathcal{P}}_{m}$ be the Cauchy densities on these cuts.
Subtracting the two integral representations of
the same function $S_{+}(z)$ one from the other,
we see that the ``old'' and
``new'' cuts combine into closed loops and the Cauchy
integrals over these loops vanish
for all $z$ outside the loops.
This means that the function $\tilde{\mathcal{P}}_{m}$ is
just an analytical continuation of the function
$\mathcal{P}_{m}$,
so $\mathcal{P}_{m}(\tau )$ should be regarded
as analytic function of the complex variable
$\tau $ defined by
analytic continuation from the contour $\Gamma_{m}$.
Therefore,
given analytic functions $\mathcal{P}_{m}(\tau )$
and branch points, the choice of
cuts $\Gamma_{m}$ (if all of them lie in $D(t)$) is irrelevant.
For the LG dynamics to be reconstructed
from the multi-cut ansate \eqref{eq:Scuts},
all this means that constants of motion
are the functions
$\mathcal{P}_{m}(z)$, the branch points $z_m$ and the
constant $t_1$. Note that $\infty$ is a branch point unless
$\sum_m \mathcal{P}_{m}(z)=0$.

In this section, we consider only Cauchy densities
that are analytically extendable without
singularities from each cut to the whole
domain $D(t)$ including
infinity. For example, one may keep in mind functions regular in
$D(t)$ including infinity with fixed singularities outside
$D(t)$.
If the Cauchy densities are allowed
to have singularities in $D(t)$, then the analysis
becomes substantially
more complicated.
In the next section, we extend the construction to
the simplest possible singularity of the $\mathcal{P}_{m}$,
a simple pole at infinity. This class of solutions includes
the important case of {\it linear densities}.

It is clear that for Cauchy densities which are constant
in the physical plane,
$\mathcal{P}_m(z)=2\pi i \bar A_m$,
the function $S_{+}$ is a linear combination
of logarithms.
So, the logarithmic solutions form a
subset of the general multi-cut ones.
On the other hand, the integral representation \eqref{eq:Scuts}
suggests that
the multi-cut case can be formally thought of as a
limit of either rational or (after integrating
by parts in \eqref{eq:Scuts}) multi-logarithmic one,
with infinitely many
singularities forming a dense set of points
concentrated along contours. This remark
might be useful for the physical interpretation of
the multi-cut solutions.

\subsection{An Integral Equation for the Conformal Map}
\label{sec:BCC}

The general form of the integral equation obeyed by the
conformal map $f(w)$ is \eqref{eq:int1} (or \eqref{eq:int4}).
Similarly to the rational and logarithmic examples,
it can be simplified by shrinking the integration contour
to the cuts.
Namely, plugging
$$
\bar S_{+}(z)=\bar t_1 +
\sum_{m}\int_{\bar z_{m}}^{\infty}
\frac{z\bar{\mathcal{P}}_{m}(\tau )}{\tau (\tau -z)}
\frac{d\tau }{2\pi i}
$$
into the r.h.s. of \eqref{eq:int1}, we get
\begin{eqnarray}
&&f(w) =
rw+u_0 + \nonumber\\
&&\frac{1}{2\pi i} \sum_m
\oint_{|\zeta |=1}\frac{d\zeta}{w-\zeta}
\int_{\bar z_{m}}^{\infty}
\frac{\bar f(1/\zeta )
\bar{\mathcal{P}}_{m}(\tau ) }{\tau (\tau -\bar f(1/\zeta ))}
\frac{d\tau }{2\pi i}\nonumber
\\
&& = rw+u_0 - \nonumber\\
&&\frac{1}{(2\pi i)^2} \sum_m
\int_{\bar z_{m}}^{\infty}
\frac{d\tau \bar{\mathcal{P}}_{m}(\tau )}{\tau}
\oint_{|\zeta |=1}
\frac{\bar f(1/\zeta ) \, d\zeta}{(w-\zeta )(\bar f(1/\zeta )-\tau )}.\nonumber
\end{eqnarray}
Since $\tau$ lies in the conformal region
$D(t)$, there is a unique $\zeta_{*}$
inside the unit disk such that $\bar f(1/\zeta _{*} )=\tau$.
The last integral can be calculated by taking the residue
at the $\zeta_{*}$ (recall that $w$ is outside).
After that it is convenient to change the integration
variable $\tau \to \xi$ connected with $\tau$ by the
relation $\bar f(1/\xi )=\tau$.
As a result, we obtain the following integral equation
for the conformal map \footnote{Strictly
speaking, it is a system of two integral equations
for two functions, $f(w)$ and $\bar f(w)$.}:
\begin{equation}\label{eq:fcutsP}
f(w) = rw + u_0 -
\sum_{m}\int_{0, \gamma_m}^{a_{m}}
\frac{\bar{\mathcal{P}}_{m}(\bar{f}(1/\xi ))}{\xi - w}
\frac{d\xi}{2\pi i}\,.
\end{equation}
Here the points $a_m$ are such that $z_m =f(1/\bar a_m )$.
The integration contours $\gamma_m$ are determined by the
change of the integration variable. Specifically, consider a contour
$\gamma_{m}^{*}$ from $\infty$ to $1/\bar a_m$ in the mathematical
plane, which is the pre-image of $\Gamma_m$ under the map $f$:
$\Gamma_{m}=f(\gamma^{*}_{m})$, then $\gamma_m$ is the contour
connecting $0$ and $a_m$ obtained as the inversion of the
$\gamma^{*}_{m}$ with respect to the unit circle (the transformation
$w \to 1/\bar w$).
All the parameters in \eqref{eq:fcutsP} except for coefficients of the
functions $\bar{\mathcal{P}}_{m}(z)$ depend on $t$. The integration
contours depend on $t$ as well. However, as is explained in the
previous subsection, the integral does not depend on a
particular shape of the contours provided they do not intersect
each other and the boundary of the domain $D(t)$. In what follows
we do not indicate the integration contours explicitly.
Note that equation \eqref{eq:fcutsP} can be analytically
continued inside the unit disk provided $w$ does not
intersect the cuts from $0$ to $a_m$.

In the logarithmic case, $\mathcal{P}_{m}(z)=2\pi i\bar A_m$,
the r.h.s. of equation \eqref{eq:fcutsP}
immediately gives the logarithmic ansate for $f$
with moving branch points and constant coefficients.
In fact it is the integral equation \eqref{eq:fcutsP} that
justifies the logarithmic ansate.

Because of importance of the integral equation \eqref{eq:fcutsP}
we give here an alternative derivation ``by hands" which is longer but
less formal and probably more instructive.
Using the same arguments as in Section
\ref{sec:polsAndLogs}, one can see
that for each
branch point $z_m$ of the function
$S_{+}(z)$ there is
a corresponding time-dependent branch point $a_m$
of the function $f(w)$ determined by the relation
$z_{m}=f(1/\bar a_{m})$.
The conformal map can then be written as
\begin{equation}\label{eq:fcuts}
f(w) = rw + u_0 + \sum_{m}\int_{0,\gamma_m}^{a_{m}}
\frac{\rho_{m}(\zeta)}{\zeta- w}\frac{d\zeta}{2\pi i},
\end{equation}
where $\rho_{m}(\zeta )$ are time-dependent
Cauchy densities on the
cuts $\gamma_{m}$ between $0$ and $a _{m}$.

\begin{figure}
\includegraphics[width=\FigWidth]{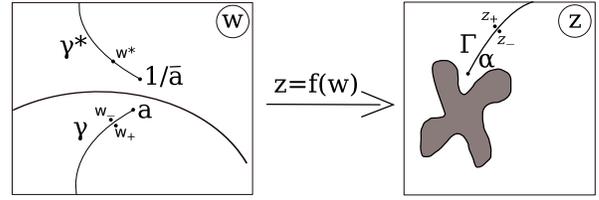}
\caption{\label{fig:cuts}
The correspondence between cuts in the mathematical
and physical plane.
}
\end{figure}

Let us fix a path $\Gamma_{m}$ from $\infty$ to
$z_{m}$ lying entirely inside the domain
$D(t)$ in the physical plane.
It is an image of a path $\gamma^{*}_{m}$
connecting the points $\infty$ and $1/\bar a_m$
outside the unit disk: $\Gamma_{m}=f(\gamma^{*}_{m})$.
Let us reflect $\gamma^{*}_{m}$ with respect to the unit circle
(i.e., make the transformation $w \to 1/\bar w$).
The reflected path, $\gamma_{m}$, connects the points
$0$ and $a_{m}$ inside the unit disk. We take these
$\gamma_{m}$ to be
the integration paths in \eqref{eq:fcuts}.

Take a point $w\in \gamma_{m}$ and
consider two points, $w_{+}$ and $w_{-}$, which are limits
to $w$ from
respectively left and right sides of the
branch cut $\gamma_{m}$ (oriented from
$0$ to $a_m$).
Similarly to \eqref{eq:jump1}, we have $f(w_+)-f(w_-)=
\rho_m (w)$. The corresponding points in the physical plane,
$z_{\pm}$, tend from opposite sides
to the point $z=f(w^*)\in \Gamma_m$,
where $w^* =1/\bar w \in \gamma^{*}_{m}$ is the image of $w$ under
the inversion (see Fig. \ref{fig:cuts}).
Since conformal maps preserve
orientation, we can write $z_{\pm}=f(w^{*}_{\pm})$.
However, the inversion interchanges the sides, i.e.,
$w^{*}_{\pm}=1/\bar w_{\mp}$, so $z_{\pm}=f(1/\bar w_{\mp})$.
According to equations \eqref{eq:fS}
rewritten as
\begin{equation}
\label{eq:fS1}
\left\{
\begin{array}{l}
S = \bar f(\bar w) \\
z = f(1/\bar w)\,,
\end{array}
\right.
\end{equation}
the values of the Schwarz
function at $z_{\pm}$
are $S(z_{\pm})=\bar{f}(\bar{w}_{\mp})$ and
the discontinuity of the Schwarz function
is thus $[S(z)]=S(z_{+})-S(z_{-})=
\bar{f}(\bar{w}_{-})-\bar{f}(\bar{w}_{+})=
-\bar{\rho_{m}}(\bar{w})$. As singularities of
$S_{+}$ and $S$ in $D(t)$ are the same,
this is exactly the discontinuity of the
function $S_{+}(z)$. We thus conclude that
on $\gamma_m$ it holds
\begin{equation}
\label{eq:rho}
\rho_{m}(w)=-\bar{\mathcal{P}}_{m}(z), \quad
z=\bar{f}(1/w)
\end{equation}
Furthermore, since the r.h.s. is analytically
extendable to the whole $\bar D(t)$
and the function $\bar{f}(1/w)$ sets up
a conformal equivalence between $\bar D(t)$ and the
unit disk, the $\rho_{m}(w)$ is analytically extendable
from the cut $\gamma_m$ to the whole unit disk.
Therefore, \eqref{eq:rho} actually holds
everywhere in the unit disk.
Plugging it into \eqref{eq:fcuts}, we obtain
the non-linear integral equation \eqref{eq:fcutsP} for $f(w,t)$.

\subsection{The Solution Scheme}
\label{sec:scheme}

To summarize, we have the following formulas
which express constants of motion through the time-dependent
conformal map:
\beq \label{eq:multiZ}
\begin{array}{l}
\displaystyle{t_1 =
\frac{1}{2\pi i}\oint_{|w|=1}\frac{\bar f(1/w)d f(w)}{f(w)}}\,,
\\   \\
z_m \, =  f(1/\bar a_m (t) , t)\,.
\end{array}
\eeq
They look exactly like the corresponding formulas
\eqref{eq:logS+1} for
the logarithmic case.
The main difference is that in general any
explicit representation of the function $f$
is not a priori available. It is defined implicitly
via the integral equation
\eqref{eq:fcutsP}.

In principle, these equations supplemented by
a formula for $t$ below give a solution to the problem
in an implicit form,
like in rational or logarithmic cases.
However, the general multi-cut construction is
substantially more complicated
because before applying the scheme outlined
in section \ref{sec:polsAndLogs} one has to find a solution
to the integral equation \eqref{eq:fcutsP} with
required analytic properties.
Clearly, such a solution depends on $a_m$, $u_0$ and $r$ as
parameters: $f(w)=f(w|r, u_0 , \{a_m\})$.
Then \eqref{eq:multiZ} is a system of equations
to determine them through the constants
of motion.
The missing equation which includes $t$
can be obtained from \eqref{eq:int5} or \eqref{eq:area1}.
One of its forms is
\beq
\label{eq:area5ab}
t=r^2 -\frac{1}{2\pi i}\sum_m \int_{0}^{a_m}
\bar{\mathcal{P}}_{m}(\bar f (1/\zeta ))\,
d (\bar f (1/\zeta )-r/\zeta )\,,
\eeq
Another form is
\beq
\label{eq:area5}
t=r^2 - \sum_{l,m}
\int_{0}^{a_l }\frac{d\zeta}{2\pi i}
\int_{0}^{a_m }\frac{d\bar w}{2\pi i}
\frac{\rho_l (\zeta )\overline{\rho_m (w)}}{(1-\zeta \bar w)^2}
\eeq
where $\rho_m (\zeta )=-\mathcal{P}_{m}(\bar f(1/\zeta ))$.
It is an analog of \eqref{eq:logarea0} (compare also with \eqref{eq:area3}).
Meanwhile, a formula for $t_1$ similar to \eqref{eq:area5ab} also exists:
\beq \label{eq:logS+1bis}
\bar t_1 =u_0 -\frac{1}{2\pi i}\sum_m \int_{0}^{a_m}
\bar{\mathcal{P}}_{m}(\bar f (1/\zeta ))\,
d \log (\zeta \bar f (1/\zeta ))\,,
\eeq
which follows from \eqref{eq:int2} at $j=0$.
In the case of constant densities, formulae \eqref{eq:area5ab} and
\eqref{eq:logS+1bis} immediately give expressions
\eqref{eq:logareabis} and \eqref{eq:logS+1} for $t$ and $t_1$
from section \ref{sec:Logs}.

\section{Multi-Cut Solutions with Linear Cauchy Densities}
\label{sec:multi-linear}

The aim of this section is to elaborate the important case
of linear Cauchy densities of the Schwarz function in the physical plane.
This class of solutions is
characterized by the property that the second derivative
of $S_+$ is a rational function (or, equivalently,
the second derivative of $S$ is a mesomorphic function in
$D(t)$). The integral equation for conformal map which is in general non-linear,
becomes linear in this case.
In the context of the
universal Whitham hierarchy, solutions of this type
were discussed in \cite{Krichever94} (section 7.2) and
in \cite{KMZ05} (section 3.8). This class
also includes logarithms and poles
on the background of the multi-cut functions.
However, the approach of the previous section is not
directly applicable because the integrals become divergent.

These divergences are artificial and can be curbed for the
price of introducing one more time-dependent parameter.
The integral equation should be modified.
We start, in subsection \ref{sec:simplepole},
with a more general situation when the Cauchy densities
$\mathcal{P}_{m}(z)$
are analytic everywhere in $D(t)$ except for a
pole at infinity.
In subsection \ref{sec:linear} we specify the results to
the most important case of purely linear Cauchy densities.
In subsection \ref{sec:wedgeT} we consider, as an example, some special
``finger''  patterns with $\ZZ _N$ rotational symmetry
(which are equivalent to solutions in a wedge with angle $2\pi/N$).
Their asymptotic form at large $t$ is given by
the known family of $\ZZ _N$-symmetric self-similar ``fingers"
described by hypergeometric solutions to the integral equation
\eqref{eq:fcutsP} (see
\cite{TRHC,Benamar,BAmar,Combescot,Cummings1999,Richardson2001,MV}).

\subsection{Cauchy Densities with a Pole at Infinity}
\label{sec:simplepole}

Let us consider the case when $\mathcal{P}_{m}(z)$
are analytic everywhere in $D(t)$ except for a simple pole
at infinity.
One immediately sees that neither the multi-cut ansate \eqref{eq:Scuts}
nor the integral equation \eqref{eq:fcuts} can be directly applied to this
case because the integrals diverge. Nevertheless, it is possible to modify these
formulas in such a way that the divergences disappear. To this end, consider
a modified multi-cut ansate:
\beq \label{eq:modif1}
S_{+}(z)= t_1 + 2t_2 z + \sum_m \int_{\infty , \Gamma _m}^{z_m}
\frac{z^2 \mathcal{P}_{m}(\tau )}{\tau^2 (\tau -z)} \, \frac{d\tau}{2\pi i}\,,
\eeq
where $t_1$ and $t_2$ are arbitrary complex constants (the first and the
second harmonic moments) and
we adopt the same conventions about the cuts
$\Gamma_m$ as before. A single-valued branch of this function
is fixed by the conditions $S_{+}(0)=t_1$, $S_{+}'(0)=2t_2$.
Let us plug it
into the integral equation \eqref{eq:int4a}
with $d=1$. A simple calculation similar to the one done in section
\ref{sec:BCC} yields the integral equation
\beq \label{eq:modif2}
f(w)=rw + u_0 +\frac{u_1}{w}-  \sum_m
\int_{0}^{a_m} \frac{\zeta \bar{\mathcal{P}}_{m}
(\bar f(1/\zeta ))}{w(\zeta -w)}\,
\frac{d\zeta}{2\pi i}
\eeq
The general scheme of solution remains the same, with the
only difference that now we have one more time-dependent parameter
(the coefficient $u_1$). Accordingly, we need an extra equation
connecting it with integrals of motion. The necessary equations,
which generalize \eqref{eq:logareabis} and \eqref{eq:logS+1},
can be obtained
from \eqref{eq:int2}, \eqref{eq:int3}. They are:

\begin{eqnarray}\label{eq:X1}
&&2\bar t_2 = \frac{u_1}{r}+\frac{1}{2\pi i}\sum_m
\int_{0}^{a_m}\bar{\mathcal{P}}_{m}(\bar f(1/\zeta ))
d\left (\frac{1}{\bar f(1/\zeta )} - \frac{\zeta}{r}\right )
\nonumber
 \\
&&\bar t_1  = u_0 -\frac{u_1 \bar u_0}{r}-\nonumber\\
&&\frac{1}{2\pi i}\sum_m
\int_{0}^{a_m}\bar{\mathcal{P}}_{m}(\bar f(1/\zeta ))
d\left ( \log (\zeta \bar f(1/\zeta )) -\frac{\bar u_0 \zeta}{r} \right )
\nonumber
\\
&&t=r^2 -|u_1|^2 -\nonumber\\
&&\frac{1}{2\pi i}\sum_m
\int_{0}^{a_m}\bar{\mathcal{P}}_{m}(\bar f(1/\zeta ))\,
d\left (\bar f(1/\zeta )-\frac{r}{\zeta}-\bar u_1 \zeta \right ).
\end{eqnarray}

Let us briefly comment on a more general case of Cauchy densities
with a higher pole at infinity (and analytic
everywhere else in $D(t)$). It should be already clear how to proceed.
If the leading term of $\mathcal{P}_{m}(z)$
as $z \to \infty$ is $z^d$, then one should extract from
$S_{+}(z)$ a polynomial
$\sum_{k=1}^{d+1}kt_kz^{k-1}$ of degree $d$ and represent the
remaining part (which is of order $O(z^{d+1})$ as $z\to 0$) as
integrals along the cuts. Plugging this
ansate into \eqref{eq:int4a}, one obtains an integral equation
for the conformal map containing a Laurels polynomial
$rw + \sum_{j=0}^{d}u_j w^{-j}$ in the r.h.s. So, apart from
positions of branch points, there are $d+2$ time-dependent
parameters $r, u_0 , \ldots , u_d$ which are to be connected
with integrals of motion by formulae similar to \eqref{eq:X1}.

\subsection{Linear Cauchy Densities}
\label{sec:linear}

The case of linear homogeneous Cauchy densities is especially important.
Set
\beq \label{eq:lin1}
\mathcal{P}_m (z)=2\pi i \, c_m z\,,
\eeq
where $c_m$ are arbitrary complex constants. The explicit form
of the function $S_+$ is:
\beq \label{eq:lin2}
S_+(z)=t_1 +2 t_2 z + z\sum_m c_m \log \left (
1-\frac{z}{z_m}\right )\,,
\eeq
It is clear that the integral equation \eqref{eq:modif2} becomes linear:
\beq \label{eq:inf1}
f(w)=rw + u_0 +\frac{u_1}{w}+\sum_m \bar c_m \int_{0}^{a_m}
\frac{\zeta \bar f(1/\zeta ) d\zeta}{w\, (\zeta -w)}\,.
\eeq
Formulae \eqref{eq:pole1} read:
\begin{eqnarray}\label{eq:X2}
&&2\bar t_2 = \frac{u_1}{r}-\sum_m \bar c_m
\int_{0}^{a_m}\bar f(1/\zeta )\,
d\left (\frac{1}{\bar f(1/\zeta )} - \frac{\zeta}{r}\right )
\nonumber \\
&&\bar t_1  = u_0 -\frac{u_1 \bar u_0}{r}+\nonumber\\
&&\sum_m \bar c_m
\int_{0}^{a_m}\bar f(1/\zeta )\,
d\left ( \log (\zeta \bar f(1/\zeta )) -\frac{\bar u_0 \zeta}{r} \right )
\nonumber \\
&&t=r^2 -|u_1|^2 +\nonumber\\
&&\sum_m \bar c_m
\int_{0}^{a_m}\bar f(1/\zeta )\,
d\left (\bar f(1/\zeta )-\frac{r}{\zeta}-\bar u_1 \zeta \right )
\end{eqnarray}
After some simple transformations, they can be brought to the form
\begin{eqnarray}
&&2\bar t_2 =\frac{u_1}{r}+\nonumber\\
&&\sum_m \bar c_m \left [
\log \left (\frac{a_m \bar z_m}{r}\right )+ \frac{1}{r}
\int_{0}^{a_m}\left (\bar f(1/\zeta )-r/\zeta \right )d\zeta \right ]
\label{eq:X3}\\
&&\bar t_1 =u_0 -2\bar u_0 \bar t_2 +\nonumber\\
&&\sum_m \bar c_m \left [\bar z_m +
\bar u_0 \log \left (\frac{a_m \bar z_m}{r}\right )
-\bar u_0-\frac{r}{a_m}+\right.\nonumber\\
&&\left.
\int_{0}^{a_m}\left (\bar f(1/\zeta )-r/\zeta -\bar u_0\right )
\frac{d\zeta}{\zeta}\right ]\label{eq:X4}\\
&&t = r^2 -2r \bar u_1 \bar t_2+\nonumber\\
&&\sum_m \bar c_m \left [ \frac{\bar z_{m}^{2}}{2}+r\bar u_1
\log \left (\frac{a_m \bar z_m}{r}\right )-\frac{\bar u_{0}^{2}}{2}
-r\bar u_1 -\frac{r^2}{2a_{m}^{2}}-\right.
\nonumber \\
&&\left.\frac{r\bar u_0}{a_m}
+r\int_{0}^{a_m}\left (\bar f(1/\zeta )-r/\zeta -\bar u_0
-\bar u_1 \zeta \right )
\frac{d\zeta}{\zeta ^2}\right ]\label{eq:X5}
\end{eqnarray}

Let us specify the scheme of solution to the case
of linear densities.
Suppose one is able to find a solution to the linear
integral equation \eqref{eq:inf1} depending
on the parameters $a_m$, $u_0$, $u_1$ and $r$:
$f(w)=f(w|r, u_0 , u_1, \{a_m\})$ (with fixed $c_m$).
Then, since $f(1/\bar a_m)=z_m$
are constants of motion, we get a system of equations
for $a_m$, $u_0$, $u_1$ and $r$ which becomes closed
after adding equations \eqref{eq:X3}, \eqref{eq:X4},
\eqref{eq:X5}.
This system determines the parameters as implicit functions
of $t$.

We conclude the subsection with a remark that the solutions of the
linear integral equation \eqref{eq:inf1} for the conformal map
$f(w,t)$, which correspond to linear Cauchy density of the Schwarz
function defined by \eqref{eq:lin2}, deserve a special name in view
of their importance.  So, in what follows we will refer to them as
to {\it hyper-logarithmic solutions}, since they contain {\it
logarithmic} solutions \eqref{eq:logF} and the Gauss {\it
hyper}-geometric function (see \eqref{confmap} below) as particular
cases.

\subsection{Example: $\ZZ _{N}$-Symmetric Solutions with
$N$ Radial Cuts}\label{sec:wedgeT}

In this subsection we consider perhaps the simplest
non-trivial application of the linear integral equation
derived above: growing patterns with $\ZZ _{N}$ rotational
symmetry whose Schwarz function has exactly $N$ radial branch cuts
in $D(t)$ with linear Cauchy densities.

Let $\omega =e^{2\pi i/N}$ be the primitive root
of unity of degree $N$. The $\ZZ _{N}$-symmetry implies
$S(\omega z)=\omega^{-1} S(z)$. Moreover, this relation holds
for both $+$ and $-$ parts of the Schwarz function separately:
$S_{\pm}(\omega z)=\omega^{-1} S_{\pm}(z)$.
This prompts us to
choose the constants of motion in the form
\beq \label{eq:ini1}
\!\!\!c_m = c\omega ^{-2m+2}, \quad z_m =\lambda \omega^{m-1},
\quad m=1,2, \ldots , N,
\eeq
with real positive constants $c$ and $\lambda$,
so the function $S_{+}$ is
\begin{eqnarray} \label{eq:ini1a}
&&S_{+}(z)=cz\sum_{m=0}^{N-1}\omega^{-2m}\log
\left (1-\frac{z\omega^{-m+1}}{\lambda}\right )
=\nonumber\\
&&-\frac{cz^{N-1}}{(1-\frac{2}{N}) \lambda^{N-2}}
\! \phantom{a}_{2}F_{1}
\left (\begin{array}{c}1 , \,\, 1 \! -\! \frac{2}{N} \\
2-\frac{2}{N} \end{array}, (z/\lambda )^{N} \right ),
\end{eqnarray}
where $\!\!\! \phantom{a}_{2}F_{1}$ is the Gauss hypergeometric function.

For the conformal map the $\ZZ _{N}$-symmetry means
$f(\omega^k w)=\omega^k f(w)$, so the expansion of the function
$f(w)/w$ goes in powers of $w^{-N}$:
$$
f(w)/(rw) = 1+\sum_{k\geq 1}f_{k}w^{-kN}
$$
where $f_k \equiv u_{kN}/r$.
The reality of $c$ and $\lambda$ in \eqref{eq:ini1a} implies that
the coefficients $f_{k}$ are real, i.e.,
$\bar f(w)=f(w)$. In what follows we assume that $N\geq 3$,
so the coefficients $u_0$ and $u_1$ in the expansion of
$f(w)$ \eqref{eq:norm} are always identically zero.

We use the integral equation \eqref{eq:inf1}.
Substituting our data, we get
\beq \label{eq:inf5}
f(w)=rw  +
c \sum_{m=0}^{N-1}\omega^{2m} \int_{0}^{a \omega^{m}}
\frac{\zeta  f (1/\zeta )d\zeta}{w(\zeta -w)}
\eeq
where $a$ is a point between $0$ and $1$ such that
$f(1/a )=\lambda$. Choosing the integration paths to be
straight lines from $0$ to $a \omega^m$ and using the
$\ZZ _{N}$-symmetry, we arrive at the equation
\beq\label{eq:fT}
f(w)=rw + c N \int_{1/a}^{\infty}
\frac{w f(x)\, dx}{1-w^N x^N},
\eeq
where $c$ is a constant and $r$, $a$ depend on time.
Let $f(w)=f(w|r, a )$ be a solution to this
integral equation with parameters $r, a$,
then three parameters $r$, $a$ and $t$ are connected
by two equations which determine $r$, $a$ as implicit
functions of $t$. The first of these equations
is obtained from the fact that
$\lambda = f(1/a (t), t )$ is a constant of motion by
substituting $w=1/a $ into \eqref{eq:fT}:
\begin{equation}\label{eq:rAlpha}
r=\lambda a -cN \int_{1/a}^{\infty}
\frac{f(x|r, a )\, dx}{1- x^{N} a^{-N}}\,,
\end{equation}
the second is one or
another version of the formula for $t$ in terms of $f(w)$.

\begin{figure}
\includegraphics[width=\FigWidth]{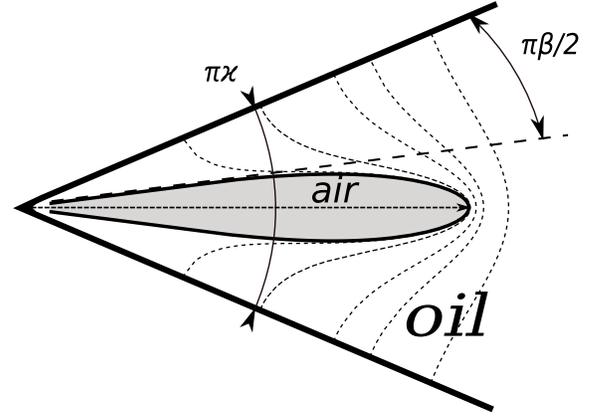}
\caption{\label{fig:wedge}
Laplacian Growth in wedge geometry. Air is pushed into oil trough the
corner of the wedge. $\beta $ is the fjord angle.
}
\end{figure}

We have obtained a one-parametric family of exact solutions
to the LG describing growth of $N$ symmetric air fingers or,
what is mathematically the same, growth of an air finger in the
wedge with interior angle $2\pi / N$ (Fig. \ref{fig:wedge}).
It is not clear at the moment whether the conformal map $f(w)$
is available in a closed analytic form. However, the solution can
be analyzed numerically (see appendix \ref{sec:numerical}).

What is the meaning of the parameter $c$? To answer this question,
consider the limit $t \to \infty$, wherein, as one
can easily show, $a \to 1$. Then equation \eqref{eq:fT}
becomes identical to the linear integral equation describing
self-similar growth of $\ZZ _N$-symmetric fingers
(or fingers in the wedge with interior angle $2\pi /N$):
\beq\label{eq:SZ6}
f(w)=rw +\frac{N\sin \pi \beta }{\pi}\int_{1}^{\infty}
\frac{w f(x)\, dx}{1-w^N x^N}
\eeq
with the ``boundary condition" $f(1)=0$.
As shown in our earlier paper \cite{AMZ}, $\beta$
is the interior angle of the oil fjord
between two neighboring air fingers.
Since $c$ in \eqref{eq:fT} is a constant of
motion, we conclude that setting
\beq \label{eq:self1}
c=\frac{\sin \, \pi \beta}{\pi}
\eeq
we can interpret $\beta$ as the asymptotic fjord angle.

Meanwhile,
an analytic solution of \eqref{eq:SZ6} in terms of
the Gauss hypergeometric function is available:
\beq\label{confmap}
f(w)=rw(1-w^{-N})^{\beta}
\phantom{a}_{2}F_{1} \left (
\begin{array}{c}\beta , \,\, \beta \! -\! \kappa \\
1-\kappa \end{array}, w^{-N} \right ).
\eeq
where
$$
\kappa = \frac{2}{N}\,.
$$
For more details, see \cite{AMZ},
where the self-similar case was studied in detail.

The integral equation \eqref{eq:fT} is equivalent to
an infinite system of linear equations for the coefficients $f_k$.
To represent it in this form, let us introduce a function $F$ via
$$
f(w)=rwF(w^{-N})\,, \quad F(w)=\sum_{k=0}^{\infty} f_k w^k\,, \quad f_0 =1\,,
 $$
then the integral equation \eqref{eq:fT} becomes
\beq\label{eq:fT1}
F(w)=1 + c \int_{0}^{\alpha}
\frac{x^{-\kappa}F(x)\, dx}{x-w^{-1}}\,,
\eeq
where $\alpha = a^N$. Substituting the Taylor expansion of $F$ and
comparing the coefficients, we get
\beq\label{eq:fT2}
f_j = -c \sum_{k=0}^{\infty} \frac{\alpha^{j+k-\kappa}}{j+k-\kappa}\, f_k\,,
\quad j\geq 1\,.
\eeq
(However, because the r.h.s. contains $f_0 =1$, this is not an
eigenvalue equation but rather an inhomogeneous system of
linear equations.) It is also useful to note the formula for $t$,
\beq \label{eq:X6}
t=r^2 +\frac{1}{2}Nc\lambda^2 +cr^2
\sum_{n\geq 0}\frac{\alpha^{n-\kappa}f_n}{n-\kappa}\,,
\eeq
which follows from \eqref{eq:X5}.

\section{Discussion and Conclusion}\label{sec:discussion}

In conclusion, let us briefly comment
on the following three major
points of this work:

\begin{itemize}
\item The inverse potential problem integral equation,
\eqref{eq:int1} (or \eqref{eq:int4}),
\item The equations for multi-cut solutions,
\eqref{eq:Scuts} and \eqref{eq:fcutsP},
\item The equations for hyper-logarithmic solutions,
\eqref{eq:lin2} and \eqref{eq:inf1}.
\end{itemize}

\noindent
{\it The inverse potential problem equation} \eqref{eq:int1}:
The formulation of the LG problem as a linear growth of
a domain area with conserved harmonic moments
requires to solve the inverse potential problem,
namely to find the conformal map $f(w)$
from the function $S_+(z) = \sum_{k=1}^\infty k t_k z^{k-1}$,
whose Taylor coefficients are conserved harmonic moments of
the growing domain
and from the area/time $t$.  While
it is straightforward to obtain $S_+(z)$ and $t$ from $f(w)$
(the direct potential problem), it is a
well-known challenge to do it the other way around (the inverse potential
problem) \cite{Gust}.
The reformulation
of the inverse potential problem
as the nonlinear integral equation \eqref{eq:int1} (or \eqref{eq:int4})
for the function $f(w)$, assuming that $S_+(z)$ is known,
appears to be extremely useful.
In particular, it allows us
to obtain some important classes of
solutions to the LG problem with
prescribed integrals of
motion (harmonic moments), including those with arbitrary branch
cuts as well as those with pole singularities.
This integral equation is also expected
to be of significant help in a future work on LG and related
problems of interface dynamics.

{\it The equation for multi-cut solutions, \eqref{eq:fcutsP}}:
A mathematical motivation for this work was a strong
feeling that rational and logarithmic conformal maps
do not exhaust the list of exact solutions of LG at
zero surface tension.  A physical motivation came from a long
standing need to describe most general
moving interfaces in Hele-Shaw experiments with negligible surface tension.
As was already said in section \ref{sec:remarks}, while finite linear combinations
of logarithms can describe interface dynamics without finite time
singularities, they fail to describe an interface with non-parallel fjords walls.
This motivated us to look for a more general family of solutions to the LG equation
\eqref{eq:LGE}.  We expect that the multi-cut solutions presented
in this paper do explain the
experiments \cite{Couder,HLS-DLA,Leif} where formation and development of oil fjords
with non-parallel and/or non-straight walls was observed.

Two special kinds of branch points and cut singularities were already
considered in earlier
works on Laplacian growth.  Fractional time-independent branch
points appeared in studies
of self-similar singular interfaces \cite{BAmar,Tu,AMZ}, while logarithmic
branch cuts with
constant Cauchy densities appeared in \cite{H86,BP86,MS94,SM94,SM98}.
We have shown that
all of them are just various special cases and limits of the presented general construction.
In the light of our approach, it also becomes clear why the pole and
logarithmic solutions are
special: the Cauchy densities are particularly simple in these cases, so the integral
equation can be easily solved.  It also shows a road to obtain new
families of exact solutions.

{\it The equation for hyper-logarithmic solutions, \eqref{eq:inf1}}:
The mathematical significance
of the hyper-logarithmic solutions is that they correspond to the linear
integral equation \eqref{eq:inf1}, which ought to be much more accessible
to analytic treatment than the general nonlinear equation \eqref{eq:fcutsP}
for multi-cut solutions.  The physical importance lies in the belief that
the hyper-logarithmic solutions describe fjords with a constant opening
angle, in agreement with viscous
fingering experiments \cite{Leif}.  The interface dynamics simulated in
a wedge, shown on Figure \ref{fig:finiteTime}, seems
to confirm this belief, if to consider fjords central lines as walls of
virtual wedges in accordance with \cite{Couder}.  A thorough geometric
analysis of the corresponding interface dynamics based on the integral
equation \eqref{eq:inf1} will be published elsewhere.

\section{Acknowledgement}\label{sec:acknowledgement}

Ar.A is grateful to Welch Foundation for the partial support.
His work was also partially supported by NF PhD-0757992 grant.
All authors gratefully acknowledge a significant help from the
project 20070483ER at the LDRD programs of LANL: the work of
M.M-W. on this problem was fully supported by this project,
while two other authors were partially supported by the same
grant during their visits to LANL in 2008.
The work of A.Z. was also partially supported by grants RFBR
08-02-00287, RFBR-06-01-92054-$\mbox{CE}_{a}$, Nsh-3035.2008.2
and NWO 047.017.015.
A.Z. thanks the Galileo
Galilei Institute for Theoretical Physics for the hospitality
and the INFIN for partial support during the completion of this work.

\appendix

\section{Numerical solution of equation (\ref{eq:fT}) }
\label{sec:numerical}

\begin{figure}
\includegraphics[width=\columnwidth]{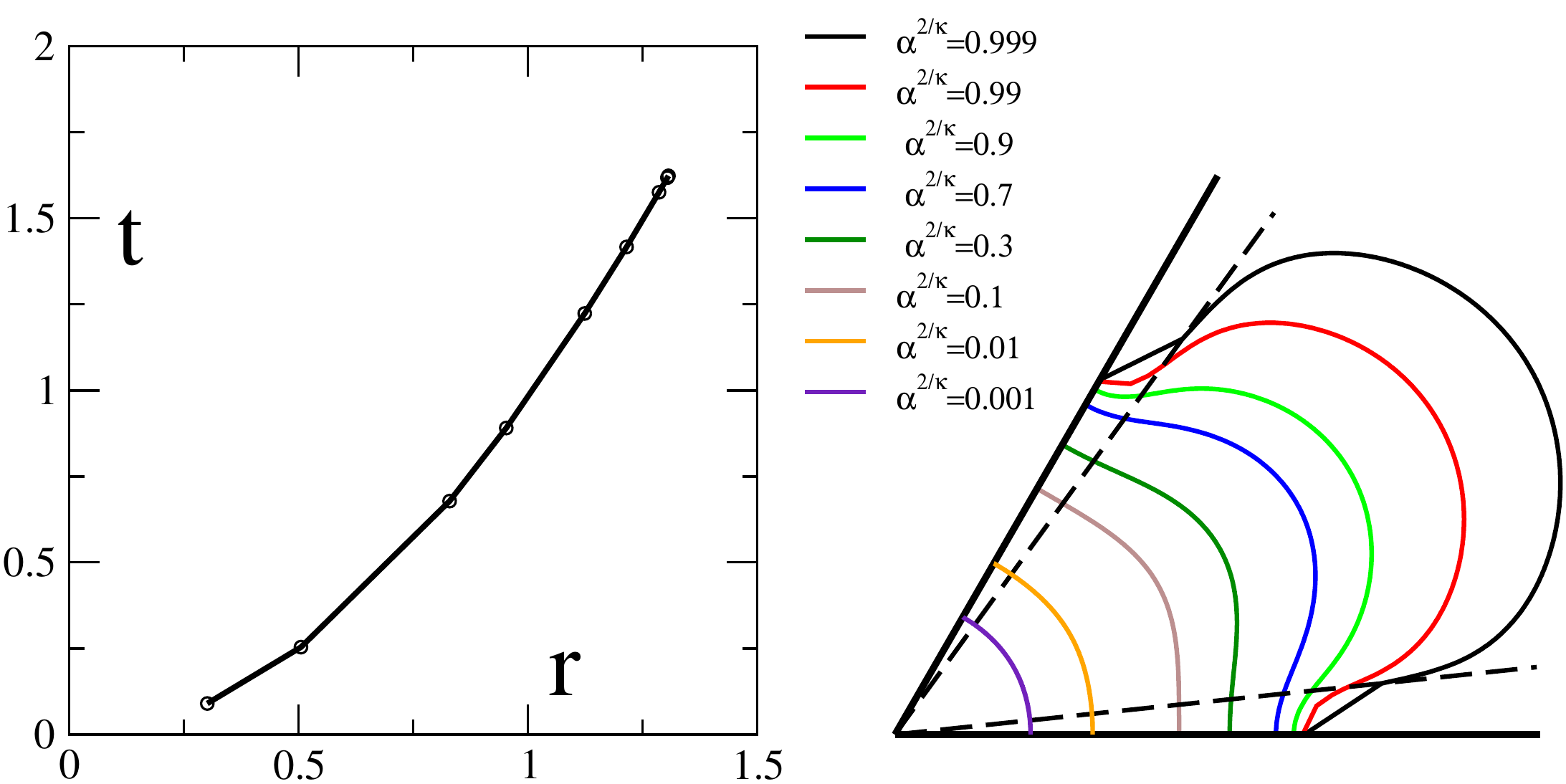}
\caption{\label{fig:finiteTime} (color online)
The results of the numerical solution of the equation \eqref{eq:fT} for
$N=6$, $\beta =0.4/N $. The left panel shows the dependence
of the area
$t$ on the conformal radius $r$, the right panel shows the growing finger
at different times.
}
\end{figure}

Equation \eqref{eq:fT}
with $c$ given by \eqref{eq:self1} can be rewritten in
the following way:
\begin{eqnarray}\label{eq:cutLogs}
&&\!\!\!\!f(w,t)\!=\!r(t)w\!-\!\frac{N\sin \pi \beta}{\pi }w
\oint_{|\zeta |=1 }\!\!\!\!\!\!\!\!\!
\frac{f(\zeta ,t)\log (1 -a(t) \zeta )}
{1 -w^{N}\zeta^{N}}\frac{d\zeta}{2\pi i},\nonumber\\
&&\!\!\!\!f(1/a(t),t )=\lambda,
\end{eqnarray}
We are looking for a
solution consistent with the $\ZZ_{N}$-symmetry:
\beq\label{eq:f1}
f(w,t)=r(t)w+\sum_{k=1}^{\infty }f_{k}(t)w^{-Nk+1}
\end{equation}
Equation \eqref{eq:cutLogs} then takes the form
\begin{eqnarray}\label{eq:fn}
&&f_{n}(t)=-r(t)\frac{\sin \pi \beta}{\pi }\frac{\alpha^{n-\kappa}(t)}{n-\kappa}
-\nonumber\\
&&\frac{N\sin \pi \beta}{\pi }\sum_{k=1}^{\infty }f_{k}(t)
\frac{\alpha^{n+k-\kappa}(t)}{n+k-\kappa}.
\end{eqnarray}
where $\kappa =2/N$, $\alpha =a^N$. Using  the relation
$\lambda =f(1/a(t),t )$ and re-denoting $f_{k}(\alpha )
\equiv f_{k}(t(\alpha ))$, $r(\alpha )\equiv r(t(\alpha ))$,
we get
\begin{equation}\label{eq:r}
r(\alpha )=\alpha^{\kappa /2}
\lambda -\sum_{k=1}^{\infty}f_{k}(\alpha )\alpha^{k},
\end{equation}
and
\begin{eqnarray}\label{eq:b}
\sum_{k=1}^{\infty}f_{k}(\alpha )\frac{k\alpha^{n+k}}{n+k-\kappa}\!\!-\!\!
\frac{\pi (n-\kappa) }{\sin \pi \beta }\alpha^{\kappa}f_{n}(\alpha )=
\lambda \alpha^{n +\kappa /2}
\end{eqnarray}
This equation is easily
solved numerically for any given $\alpha $ .
The conformal radius $r$ and the time are then found from
\eqref{eq:r} and \eqref{eq:area1} (or \eqref{eq:X6})
respectively. The solution is shown in
figure \ref{fig:finiteTime}.

\newpage

\section*{References}

\def\cprime{$'$} \def\cprime{$'$} \def\cprime{$'$} \def\cprime{$'$}
  \def\cprime{$'$}

\end{document}